# Cleaning and Surface Properties

*M. Taborelli*
CERN, Geneva, Switzerland

**Abstract**
Principles of the precision cleaning dedicated to ultra-high vacuum applications are reviewed together with the techniques for the evaluation of surface cleanliness. Methods to verify the effectiveness of cleaning procedures are addressed. Examples are presented to illustrate the influence of packaging and storage on the recontamination of the surface after cleaning. Finally, the effect of contamination on some relevant surface properties, such as secondary electron emission and wettability, is presented. This article is an updated and shortened version of the one previously published for the CAS school on the vacuum of accelerators 2006 [1].

## 1 Introduction

In an ultra-high vacuum (UHV) system a low residual gas density can be obtained and preserved only by using constituent materials having a sufficiently low vapour pressure at the working temperature [2, 3]. For the same reason, namely to reduce residual gas pressure, the surfaces facing vacuum of all the constituent parts must be free of organic additives, oils, greases, and packaging residues, which were used, for instance, during the manufacturing process.

In an accelerator, the adverse influence of the residual gas manifests in the interaction with the particle beam and the related degradation of the beam quality and lifetime [4], in the resulting dose of radiation delivered to the machine components, and in the background noise produced in the experiments [5]. The strength of such an interaction depends on the type of the beam particles, and in general, it increases as a function of the atomic number of the residual gas atoms and molecules. Carbon-containing molecules are more harmful than, for instance, hydrogen molecules which might be outgassed from the constituent metallic parts.

There are further possible reasons to apply cleaning procedures to the constituents of the accelerator. For instance, chlorine-based lubricants should either be avoided during manufacturing or completely removed from internal and external surfaces by cleaning in order to preserve a stainless steel vacuum system from corrosion during the entire lifetime of the machine. As a further example, the surface of the metal pipes delivering the working gas of the gas ionization detectors (drift tubes, wire chambers, etc.) must be free of silicones to guarantee a constant high efficiency and gas purity. Cracking of silicone molecules present in the working gas generates an insulating silicon oxide coating on the electrodes and deterioration of the detector response [6].

Various thin films, for instance non-evaporable getter (NEG) [7] or niobium superconducting coatings, are applied to the inner walls of accelerator vacuum systems. At least the same cleanliness as for the accelerator components is required for the substrate of such thin films before deposition, in order to ensure their adhesion and functional performance. For instance, fluorine and chlorine sitting on the substrate of a NEG thin film can migrate to the surface during activation and spoil the pumping performance.

Thus, different contaminations are relevant at different sites and the definitions of cleanliness and contaminant are related to the application of the parts. It is in other words a specification, which must be based on a method of control to be applied during manufacturing or before insertion into the machine and operation. The following does not address the issues related to contamination by dust and particles and considers that all the treatments are performed in a surface treatment workshop environment, without clean room equipment. This special aspect is relevant for accelerator components, where high electric fields are present, as radiofrequency accelerating cavities.

The next section illustrates the basic concepts of cleaning and cleaning methods. In the following sections the methods of cleanliness control and packaging after cleaning will also be reviewed together with the effects of contamination on some of the surface properties related to accelerator physics. Previous reviews on cleaning for accelerator technology can be found in Refs. [1, 8, 9].

## 2  Methods of precision cleaning for UHV applications

This section presents an overview of the general principles of cleaning methods suitable for UHV components. Globally, all of the methods considered below are aimed at precision cleaning, which is defined as a process leaving less than 1 μg/cm$^2$ of contamination on the surface. No universal recipe will be given, since the commercial cleaning products evolve quickly and the most appropriate solution must be selected by considering each particular application.

### 2.1  Cleaning with solvents

In a practical description, a solution is a system where a solute, in our case the contaminant, is uniformly distributed at the molecular level in a solvent without the formation of aggregates or precipitates. A detailed description of the dissolution process in terms of thermodynamics is given in Ref. [1], general concepts on the interactions in liquids can be found in Ref. [10], and only a short summary is given in the following. Since solubility depends on the interaction between the solute and the solvent, each solvent has a greater or lesser effectiveness in dissolving a given type of contamination. Molecules which are chemically similar mix more easily together and as a consequence, polar solvents will better dissolve a polar contaminant than a non-polar one. Water can easily dissolve salts and polar molecules, but is ineffective against oils and greases.

Examples of simple solvents are alcohols (methanol, ethanol, propanol, etc.) and halogenated hydrocarbons. The result of a series of tests on the performance of solvents for UHV application including hydrofluroethers (HFE) can be found in Ref. [11]. Alcohols can be easily used in laboratory applications in small amounts. For UHV applications, it is recommended to use high purity grades to avoid residues being left on the surface. Alcohols can have van der Waals interaction with polar as well as non-polar molecules, the strength of which depends on the respective aliphatic chain length. In addition, they can form hydrogen bonds. Their main disadvantages are flammability and toxicity, so that their application in large amounts (more than a simple beaker) should comply with ATEX (equipment for potentially explosive atmospheres) regulations, including typically the use in a closed vessel system. Halogen-based solvents (trichloro-ethylene, trichloro-ethane, chloroform, freon, perchlorethylene) are only slightly polar or non-polar and are very effective in dissolving many types of greases. Some of them however are now banned for toxicity or environmental impact reasons, as their impact on ozone depletion and global warming potential and allowed emissions are strictly regulated. Perchlorethylene is tolerated in working areas when the concentration is below 20 ppm or 128 mg/m$^3$ [12] (8 h exposure). More refined solvents are based on HFE can have remarkably low surface energies (e.g., 14.5 mJ/m$^2$ compared to 72 mJ/m$^2$ for water) and wet most of the surfaces. These compounds do not have an effect on ozone depletion, but their use is progressively limited due to their important global warming potential (typically 100−1000). In any case, the halogen-based solvents can only be applied in closed cleaning plants where the vapours are continuously collected and recycled.

Recently, modified alcohols (alkoxy propanol R-O-R'-OH) were introduced for industrial cleaning. They are halogen-free solvents designed for operation in closed circuit vessels. Alkoxy propanols exhibit balanced properties between polar and non-polar components and can therefore dissolve contaminants of various natures (30−70 mJ/m$^2$). They are sometimes mixed with other hydrocarbons in the cleaning product.

After selection of the best solvent for the actual application, contamination cleaning can consist of the complete immersion of the workpiece to be cleaned in the solvent bath and agitation of either the piece or of the bath. After extraction, the workpiece is dried by letting the solvent evaporate. Clearly, the solvent must not leave on the surface any residues of itself or of the dissolved contaminants and rinsing with pure solvent may be necessary. Drying must be performed through a controlled and reproducible procedure. Evaporation of the solvent often cools the part due to the absorbed heat of evaporation and recontamination through adsorption and condensation from air can be avoided only by keeping the part warmer than the surrounding atmosphere, or by drying in an oven.

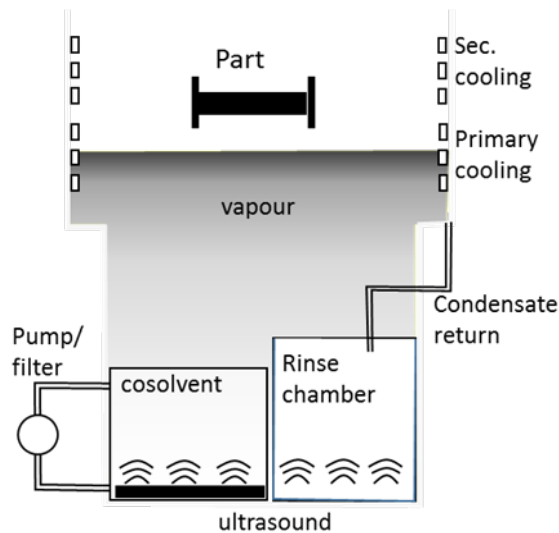

**Fig. 1:** Cleaning with co-solvent combined with vapour phase cleaning

Nowadays, the most frequent procedure occurs in a closed plant, where the solvent is recycled by re-condensing its vapour (Fig. 1). The cleaning occurs in several steps, which can include a pre-cleaning in a compatible co-solvent and a rinsing in a second solvent liquid phase, with ultrasonic agitation. The final cleaning occurs in the second solvent vapour phase. This final process does not leave, in principle, dry residues on the surface and can be applied to solvents having a sufficiently high vapour pressure slightly above room temperature. The bath of solvent is warmed up to obtain vapour above. The colder work piece is suspended in the vapour region, so that the vapour condenses onto the surface of the part to be cleaned. The condensed solvent with the dissolved contaminants will fall into the underlying bath again. The interesting aspect of this method is that the solvent is continuously distilled and only pure solvent condenses on the workpiece, whereas contamination accumulates in the bath, which can be filtered. In practice, in spite of its volatility, traces of the solvent can be detected by the residual gas analysis in the vacuum system for parts with complex shapes (edge welded bellows), which are not baked.

A special solvent is $CO_2$, which is non-polar and can dissolve aliphatic chains shorter than 20 methylene units or even silicones [13]. It is less effective for polar contaminants and residues of -C=O and COOH groups which are not eliminated from the surface and can be detected for instance by X-ray photoemission spectroscopy (XPS). In a simple procedure, $CO_2$ is used in the form of snow, where a jet of solid $CO_2$ forms a liquid layer upon impact on the surface to be cleaned [14]. During treatment the work piece should be kept above room temperature to avoid re-contamination. When used on soft substrates, such as graphite, such a treatment can induce surface roughening and therefore increase the outgassing during pump down; for this reason, tests with the respective material should be performed. In a more refined method supercritical $CO_2$ (SCCO2) is used. The gas is compressed and heated above the triple point (see dotted region in Fig. 2). The supercritical fluid can perfectly wet any surface, since it has an extremely low surface energy (1 mJ/m$^2$ for $CO_2$ compared to 32 mJ/m$^2$ for perchlorethylene) and a much lower viscosity than in its liquid phase. The solvation properties are similar to those of the liquid phase. To our knowledge, this technique is not yet available at a commercial level for the cleaning of parts for UHV. Suitable co-solvents and surfactants have been developed in order to improve the effectiveness of removing polar substances.

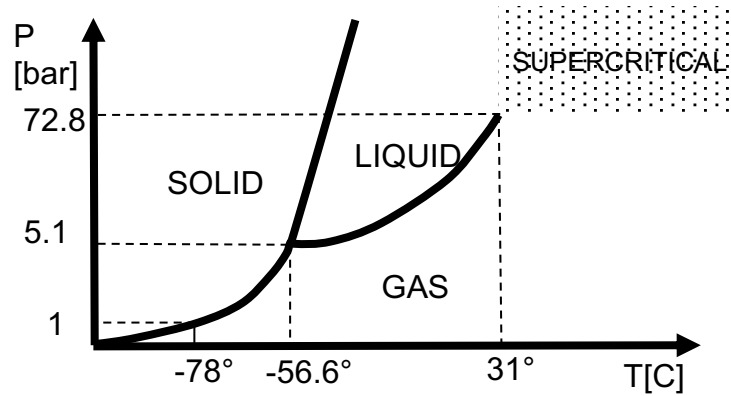

**Fig. 2:** Simplified phase diagram of $CO_2$. The supercritical region is marked with a dotted area above the critical point.

## 2.2 Cleaning with detergents

A detergent is a blend of substances designed for cleaning applications in combination with a solvent. In the following, only the most frequent case where the solvent is water will be considered. The main constituent is a surfactant (contraction of surface active agent), a substance which is able to wet virtually any surface. Such surfactant molecules are amphiphilic i.e., are able to attract both hydrophilic (through H-bonds) and hydrophobic chemical groups. They are constituted of a polar or ionic hydrophilic head group and a long hydrophobic tail, for instance an aliphatic chain. The molecule can have attractive interaction with virtually any molecules, hydrophilic or hydrophobic, polar and non-polar with the most appropriate of its ends. Therefore, such molecules like to sit at the interfaces between hydrophilic and hydrophobic media such as water−air, water−oil, etc. Above a critical concentration in the solvent (critical micelle concentration, or CMC) a surfactant can build micelles (Fig. 3). Micelles are ordered aggregates of molecules which expose their heads to water. The hydrophobic effect is the driving force to form micelles and to order the surfactant molecules. Inverse micelles can form in a hydrophobic liquid.

The cleaning mechanism is twofold. First, the surfactant removes contamination from surfaces, since it can attract many types of chemical groups and wet both the surface and the contaminant. It lowers in this way the surface energy of the substrate, avoiding re-adsorption of the contamination. The total energy is lowered already when the surfactant penetrates between the surface and the contamination, the contact area of the contaminant with the surface decreases, and removal through agitation is made easy. Second, the surfactant surrounds the contaminant, which could not otherwise be dissolved, and encapsulates it in a micelle. Recontamination is avoided and the micelle is then dispersed in the solvent. Continuous filtering of the bath removes the contamination. For instance, in water, the surfactant molecules will surround an oil droplet and encapsulate it in a micelle (Fig. 3) exposing polar heads toward water.

The detergent can contain water 'softeners', such as soluble silicates, which avoid calcium deposits. It has often a basic pH, an aspect which should be considered in view of the possible resulting surface damage (for aluminium for instance). However, generally slight etching eliminating the surface oxide, is beneficial, since contaminants can be trapped inside. The performance of the detergent is generally optimized at a given concentration and temperature (50°−60° C), which should be compatible with the parts to be cleaned. At the end of the treatment in the detergent bath, an extensive rinsing with tap water followed by demineralized water is necessary to eliminate residues of surfactant or additives, such as silicates, from the surface.

In order to guarantee a constant effectiveness of the cleaning bath over time, its quality must be periodically monitored, through pH, conductivity and concentration of detergent measurements, the latter being relevant for the formation of micelles and test specimens (see section 3). It is important to monitor the concentration of metals in the bath to avoid transfer of such ions onto other materials passing in the same cleaning station. Complicated cases in this respect are brazed joints, where one of the metals or one of the components of the brazing alloy can be spread on the entire surface of the workpiece.

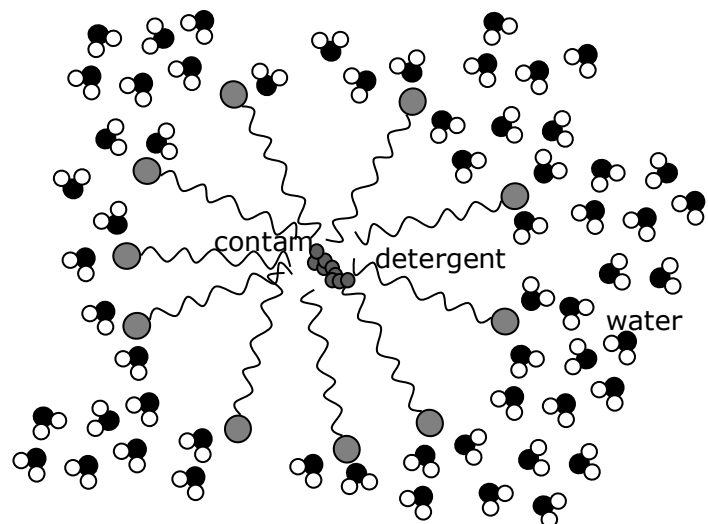

**Fig. 3:** Arrangement of surfactant molecules forming a micelle in water and encapsulating a hydrophobic contamination particle.

As described above, the cleaning power of solvents is in principle improved by detergents, which operate always in combination with a solvent. A nice comparison between various solvents and detergents used for precision cleaning of UHV components is given in Ref. [15]. The main reason why solvents are applied in some cases is the need to rinse the detergent cleaned parts with water in order to remove completely the traces of detergent from the surface. Such a rinsing can only be effective when the shape of the cleaned part does not trap water and residues through pockets, pores, and meander-like shapes. Residues can provoke long lasting outgassing or corrosion. For this reason, vapour cleaning by solvents is necessary for bellows, porous materials such as ceramics, narrow curved pipes, valves, and similar manifold components.

Moreover, since volatile solvents can be recycled by distillation during the cleaning process, they are particularly suitable for a gross cleaning phase of parts which are heavily contaminated by oils and greases after manufacturing. The gross cleaning can be used in combination with a subsequent treatment in a detergent bath. In this way, the contamination of the generally expensive detergent bath can be limited and its lifetime prolonged.

In some cases, detergents can provoke chemical deterioration of surfaces and alter their functional properties. A striking example is given by TiZrV non-evaporable getter (NEG) coatings used for pumping purposes in accelerator pipes. Cleaning by a detergent having a basic pH [1] and commonly used in standard CERN cleaning procedures for vacuum chambers results in a reduction of the surface vanadium content and an increase of silicate residues (up to 10% Si in surface analysis). As a consequence, the functional behaviour of the getter is deteriorated and no activation occurs in the useful temperature range (Fig. 4). Recent tests made with sodium hydroxide aqueous solution indicate that the problem comes probably more from the presence of the silicates on the surface than from the reduction of the vanadium content. Indeed, the activation of the NEG after treatment with sodium hydroxide was found to be satisfactory.

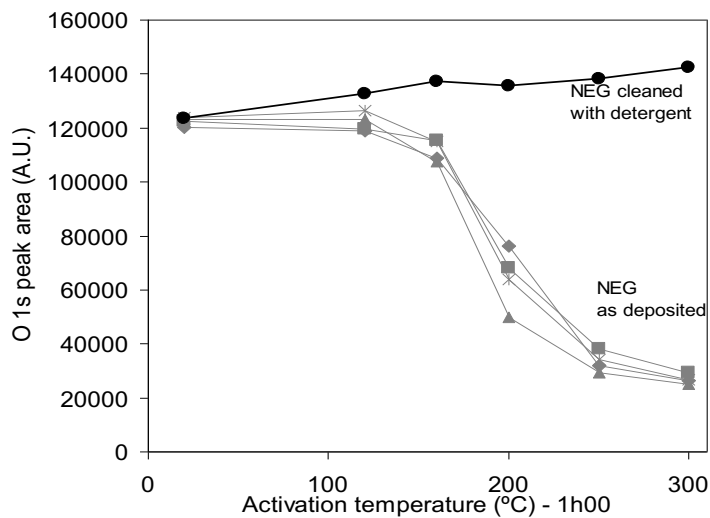

**Fig. 4:** Activation behaviour of TiZrV NEG coating illustrated by the decrease of the oxygen O1s line intensity measured in XPS. The sample cleaned in a detergent bath does not exhibit any activation.

## 2.3 Agitation

In the case of precision cleaning by immersion in a bath for both solvents and detergents, ultrasonic agitation [16] of the bath is applied. In this way, removal of soils is much more effective. Ultrasonic waves (20−120 KHz) are mechanical pressure waves generated by piezoelectric transducers placed in the cleaning tank. The waves create bubbles of some 10−100 microns in the liquid medium. Interestingly, the size of the bubbles decreases with the frequency and smaller features such as tiny blind holes for instance become accessible to the agitation. Bubbles grow up to implosion and energy is then released. In such a way, adsorbed contaminants and particles can be removed even from crevices and blind holes. It is intuitively clear that, on the one hand, smaller bubbles can carry less energy and are therefore less effective for cleaning. On the other hand, they also provoke less damages, an important aspect in case of delicate materials, such as polished copper surfaces which have undergone a thermal annealing or brazing cycle as for accelerating structures. Agitation is mandatory for all samples having a complex shape. The power of the ultrasonic actuators must be correctly dimensioned and depends on the bath volume, the position and orientation of the actuators, and the shape of the bath tank. For long pipes which cannot be easily immersed, ultrasonic agitation can be replaced by turbulent flow. The cleaning fluid is forced to stream through the pipe in the turbulent flow regime.

Examples of cleaning procedures, including some optional steps, are illustrated in the schemes in Fig. 5. The application of the various optional steps depends on the level of contamination of the parts to be cleaned, on their shape, material, application, and so on. For instance, pre-cleaning in a solvent is applied to heavily contaminated parts. The same is true for the cleaning time, which can only be defined for a specific plant and the average level of contamination of the parts to be treated.

High pressure water rinsing has been found to be successful for improving the performance of superconducting radiofrequency cavities [17], but as previously mentioned, the aspects related to dust and particle contamination are not addressed here.

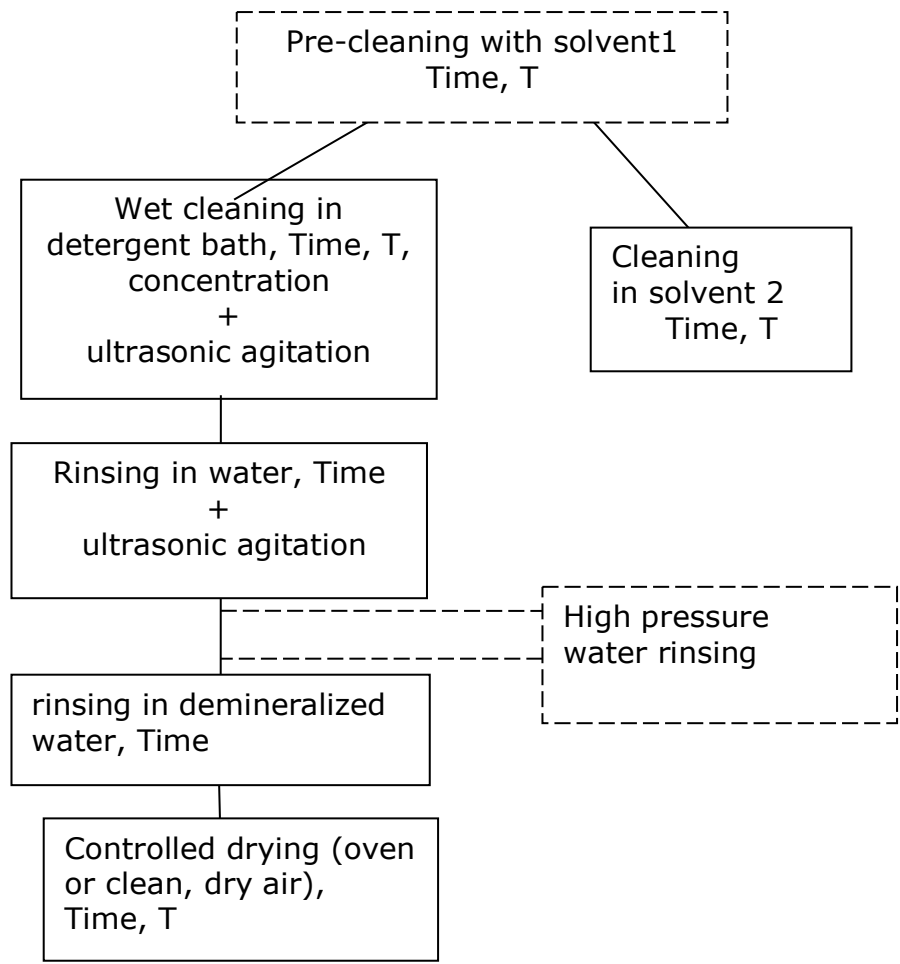

**Fig. 5:** Typical sequence for cleaning of UHV parts with solvents or detergents and typical parameters to be controlled. Dashed lines represent optional steps depending on the particular application.

## 2.4 Special cases

Vacuum greases often contain silicones, which are difficult to eliminate from the surface. Indeed, silicones can float on a cleaning bath and during extraction of the work piece from the bath the contamination again wets the surface. As already mentioned, some silicone species have non-negligible solubility in SCCO2, but no commercial system yet exists to apply this technique for cleaning of large UHV parts. An excellent solvent for many silicone greases and compounds is hexane [18]. Its effectiveness has been verified for instance by analysing by XPS a copper surface previously contaminated with silicone grease and cleaned with hexane. The silicon signal was below the detection limit. Unfortunately, hexane is toxic and highly volatile and must be used in low amounts in well vented rooms, so that the average concentration remains below 20 ppm or 72 mg/m$^3$ [19] time weighted average for 8 h. Therefore, this treatment can be applied to small parts or extraction for analytical purposes (see below).

This review considers only the cleaning methods. In addition, etching, to eliminate damaged layers or treatments to induce surface passivation and avoid, for instance, re-oxidation, can be performed if necessary. Some treatments are described in Ref. [8]. They are not suitable for parts which need to preserve accurate dimensions and precise surface finishes. Moreover, it is worth remarking that the surface in operation conditions is often not the same as after cleaning, since bake-out is frequently used for UHV systems.

## 2.5 Cleaning with ions: glow discharge

After cleaning with the methods described in sections 2.2 and 2.3, a further processing step can be applied in order to remove oxides or contaminants adsorbed upon air exposure. Glow discharge treatment can be applied to vacuum pipes after evacuation and filling with a suitable gas at low pressure ($10^{-1}$ to $10^{-3}$ mbar). Parts with other shapes can be inserted into a vacuum chamber designed for this purpose and in this case the geometry of the electrodes should be studied carefully. Glow discharge can be applied with DC, AC, or radio frequency voltages and by using various gases. The principle consists of accelerating ions toward the surface so that the surface is sputter cleaned. For long pipes, such as accelerator beam chambers, the anode is a wire placed along the chamber axis and the chamber is kept at ground potential. In general, the wire is removed after the treatment by venting and opening the chamber. If the air exposure time to remove the electrode is minimized, the beneficial effect is partly preserved [20]. Venting must be carefully performed with clean (or at least dry) gases, such as pure $N_2$. Depending on the dose of ions the treatment can remove the topmost hydrocarbon contamination or even the native oxide layer on the surface. This configuration is exploited also if additional cleaning or oxide removal is necessary prior to magnetron sputter deposition of coatings on long pipes; the anode is already in place and acts then as the cathode during the deposition process.

Sputter cleaning with a noble gas such as argon does not induce chemical reactions with the surface to be cleaned, but some implantation occurs. In stainless steel, implanted argon can be removed by baking at 350°C [21]. A mixture of gases can be used to remove more effectively a specific contaminant so that the total ion dose can be reduced. For instance, mixtures containing oxygen (Ar with 5−10% $O_2$) are very effective in the elimination of carbon from the surface [21, 22]. Hydrogen and helium are used as working gases for glow discharge treatment, especially in fusion reactor walls [23] where such gases are not harmful but are less frequently used for UHV systems. A special use of helium is the so-called helium processing performed to condition radio-frequency niobium superconducting cavities. The process is again sputter cleaning and helium is used just for convenience, since pure gas is available for the cryogenic circuit [24].

The main disadvantage of glow discharge treatment is the possible coating of insulating parts or windows with the sputtered material. The treatment can be applied to various metals such as stainless steel, aluminium, copper, titanium, and beryllium. Glow discharge on beryllium is attractive, since this material cannot be easily handled in a wet cleaning facility due to the toxicity of its oxide. In the case of beryllium, there is great advantage the fact that at, sufficiently low energies (below ~300 eV), the sputtering coefficient of $O_2^+$ on C is higher than that on beryllium thanks to a chemical reaction producing CO and $CO_2$. The effect is even more marked for beryllium oxide. Tests made at CERN with glow discharge in pure oxygen on small samples showed a strong decrease in carbon levels on the beryllium surface and no detectable (XPS) beryllium on the mounting used to hold the sample.

In other domains, glow discharge treatment is used to increase the wettability and reactivity of polymer surfaces since the plasma or sputtering breaks surface bonds.

## 2.6 Cleaning with ozone

A further method of cleaning exploits the chemical reactivity of ozone and the effect of the UV light, which is used to produce it. The mechanism of cleaning is explained in Ref. [25]. The reaction to produce ozone is induced by UV light, for instance using low pressure Hg discharge tubes delivering two main lines at 184.9 nm 253.7 nm. The 184.9 nm light is absorbed by $O_2$ in air and leads to ozone production in a two-step mechanism. First UV splits the $O_2$ molecules and then $O_3$ is built from a fragment and an oxygen molecule. The 253.7 nm light is absorbed by most hydrocarbons and by ozone and promotes their decay. The high reactive atomic oxygen can oxidize hydrocarbon species in synergy with the UV light which splits them. As shown in Ref. [25], the action of UV light alone is much less effective. This type of cleaning is therefore adapted to eliminate hydrocarbon contamination. It is not however used for a macroscopic layer of hydrocarbon such as oil or grease, since the ozone effect is only on the topmost surface of the contamination; however, the UV action penetrates deeper and could even promote polymerization in the layer. Care must be taken with oxidation effects, which are negligible on stainless steel but clearly visible on copper and aluminium (see Ref. [25] and references therein).

# 3 Cleanliness and cleaning effectiveness

## 3.1 Methods to assess cleanliness

Cleanliness must be evaluated with respect to the application for which the surface is intended. Often it is not possible to verify directly the functional performance without assembling an entire device and therefore a control technique is chosen that provides sufficient sensitivity to the known critical contaminants. The control procedure measures the amount of contaminant still present on the surface or a quantity related to it and compares this with the present limit for acceptance. This limit defines cleanliness. In general, if contamination can be quantitatively monitored one can define cleanliness classes, which are specifications for each application.

In accelerator technology various contaminants present on the vacuum chamber surface can deteriorate the vacuum (hydrocarbons, intermediate vapour pressure compounds), propagate through the system (low vapour pressure metals like Cd and Zn), promote corrosion (halogens), and/or transform into insulating layers upon irradiation (silicones). For the assessment of surface cleanliness many techniques and procedures have been used (Table 1) [26, 27]. Most of these reveal the presence of the contaminants without verifying directly the functional performance of the surface. Many require a special sample size or shape and cannot be applied to the cleaned part itself, only to a test specimen which has followed the same treatment. The techniques can be divided in two types. Analytical techniques can identify and more or less quantify the contamination. Other methods measure some quantity, which is related in a complex and sometimes obscure way to surface cleanliness, and can indicate at least whether excessive contamination is present and enable rejection or otherwise of the part before insertion in UHV.

The advantage of the analytical techniques is that, by identifying the contamination, they often enable an understanding of its origin. Common surface analyses, such as XPS, Auger electron spectroscopy (AES) and secondary ion mass spectroscopy (SIMS) are well adapted to detect and identify a broad range of organic and inorganic contaminants and have a high surface sensitivity. XPS enables easy and fast identification of the elements present on the surface and has a detection limit close to 1 atomic% (at%) at a probed depth of 1−3 nm for most elements. With a monochromatized X-ray source even distinction between organic species is possible. AES is as good as XPS from the point of view of sensitivity, but Auger lines are wider in the energy spectrum and overlap is frequent. Moreover, care should be taken when using AES to limit the current density of the electron beam, otherwise it can induce surface modifications through electron stimulated desorption (ESD) and also influence the local carbon coverage through stimulated diffusion [28]. ESD measurements indicate that desorption yields on air-exposed surfaces decrease by a factor of approximately 10 for $10^{16}$ electrons/cm$^2$. The typical primary beam currents used in Auger analysis (for instance $10^{-8}$ to $10^{-7}$A on an area of 10 × 10 μm$^2$ up to 100 × 100 μm$^2$) result in some $6 \times 10^{15}$ to $6 \times 10^{18}$ electrons/cm$^2$ in 10 seconds, which is the typical time required to acquire a full spectrum. There is experimental evidence that the carbon concentration measured by AES can depend on the impinging beam current density [28] (possibly electron gun condition) and decreases with irradiation. Therefore, the current density and the dose should be well controlled when comparing data of surface cleanliness. When using XPS, without high spatial resolution options, the amount of surface damages is reduced by a factor 10 to 100 [29]. SIMS, especially in its static version (SSIMS) and with high mass resolution, is superior from the point of view of identification of the chemical species of a contaminant and has a better sensitivity. The information is in principle more detailed, at the same time, the interpretation and quantification, which are affected by matrix effects, are not straightforward.

**Table 1:** Techniques for evaluation of surface cleanliness. Sensitivities should be taken as estimated orders of magnitude.

| Technique | On site, on cleaned pieces | Limitations (Analytical or not) | Cost | Time | Principle of measurement (detection limit for carbon species, when known) |
|---|---|---|---|---|---|
| XPS, ESCA (X-ray Photoemission Spectr.) | no | UHV, small samples (analytical) | high | slow | Electron photoemission (3% at. in the probed depth, corresponding to ~$10^{14}$ atoms/cm$^2$) |
| AES (Auger electron spectr.) | no | UHV, small samples, dose effect on carbon (analytical) | high | slow | Electron induced electron emission (~$10^{14}$ atoms/cm$^2$) |
| SSIMS (static secondary ion mass spectr.) | no | UHV, small samples, interpretation, needs calibration (analytical) | high | slow | Ion erosion coupled to mass spectroscopy ($10^{12}$ atoms/cm$^2$ [30]) |
| FTIR (Fourier transform infrared) | no | Signal depends on roughness if done directly on the surface. Very sensitive through elution method and measurement in transmissions. sensitivity, partly (analytical for organics) | high | slow | sub-monolayer of organic molecules, no sensitivity to most inorganic compounds |
| Gravimetry | no | Insufficient sensitivity | low | slow | Measure weight loss of the sample due to cleaning |
| Water contact angle | yes, for portable instruments | One model tested and found unreliable (water wettability by visual inspection generally used in workshops) | medium | Fast for portable system | Wettability of the surface by water |

| Method | | | | | |
|---|---|---|---|---|---|
| UV fluorescence | yes | No identification, not sensitive to all contaminants, needs calibration | medium | fast | Fluorescence of some organic molecules |
| UV-vis | no | Needs extraction through solvent, overlap of absorption lines (analytical) | high | slow | UV absorption, extraction method |
| OSEE (optically stimulated electron emission), SPD (surface potential difference) | yes | Does not distinguish between oxides and contaminants. Instability of reference surface | low | fast | photoelectron emission in air ($10^{15}$ molec/cm$^2$), measurement of work function due to oxidation and adsorbates ($\sim 10^{14}$-$10^{15}$ molec/cm$^2$) |
| ESD (electron stimulated desorption), PSD (photon stimulated desorption) | no | Needs suitable sample shape (partly analytical) | high | slow | Electron/Photon induced desorption of adsorbates detected by mass spectroscopy (sensitivity depends on irradiated sample size) |
| Static outgassing rate | yes | Sample shape, (partly analytical) | high | slow | Thermal desorption monitored by mass spectroscopy (depends on accumulation time) |
| TRXRF (total reflection X-ray fluorescence) | no | Needs smooth flat surface (analytical) | high | slow | X-ray fluorescence |
| Ellipsometry | no | Needs smooth flat surface | high | slow | Rotation of light polarization upon reflection ($10^{12}$ molec/cm$^2$) [30] |
| Surface tension markers | yes | Insufficient sensitivity, pieces must be re-cleaned | low | fast | Wetting by inks of various surface energies ($< 44$ mJ/m$^2$) |
| Radioactive tracer | no | Only on test sample contaminated on purpose | high | slow | Measure decrease after cleaning of radioactivity of a sample contaminated with a tracer [15] |

In the following the application of XPS is discussed more in detail. In general, the result of an XPS analysis is not the absolute concentration, such as molecules per cm$^2$ for instance. A so-called surface concentration or relative atomic concentration (at%) at the probed depth is obtained. This quantity is calculated from the measured intensities for each element, as peak areas, and the corresponding elemental calibration factors. In fact, this quantity corresponds to the true relative atomic concentration only for the case where this is uniform in the probed depth. This is rarely the case for contamination, since it is by definition on the top of the surface. However, this calculated relative atomic concentration can be used safely for the quantitative characterization of cleanliness and the comparison of results obtained with the same excitation source, the same analyser parameters and geometry (source-sample-analyser) [31].

The case of carbon, which is one of the most common contaminants, is considered more in detail as an example. It is present in hydrocarbons left by lubricants, cutting oils, rotary vane pump oils, residues of packaging materials, fingerprints, and finally airborne hydrocarbons from storage in an unprotected environment. An upper limit for the amount of carbon on the surface is therefore often adopted as the criterion for surface cleanliness, possibly accompanied by an upper limit for the total amount of other minor impurities. The validity of the so-called surface concentration, at %C, to assess cleanliness has been discussed in details in Ref. [31]. At CERN, for UHV applications a level in at %C on stainless steel is defined as the upper acceptable limit. It has been chosen based on the accumulated experience of UHV applications, showing that such a surface will have an acceptable degassing rate and will enable UHV conditions for a static vacuum to be achieved. As already mentioned, the measured at% depends on the geometry and source of the XPS instrument. The thickness of such a layer is estimated to be around 0.5 nm, assuming the common electron attenuation length values [32] and assuming a uniform layer of pure carbon to simulate the contaminant. Care should be taken when comparing cleanliness of different materials. The attenuation of the XPS signal from deeper layers, given by the attenuation length of the photoelectrons, is energy dependent. As a result, the same absolute amount of hydrocarbon on two different metals will give a different relative atomic concentration as measured by XPS. Conversion factors can be established, based on the experimental data, to compare the values measured on different materials, as described in Ref. [31]. For instance, the same amount of carbon contamination on stainless steel and copper will result in 40% at C and 44% at C, respectively.

A weakness of XPS, at least when used with a non-monochromatized X-ray source, is its incapacity to distinguish between silicones and silicates. The chemical shift of the silicon line Si $2p^{3/2}$ is similar in both cases and distinctions based on the detected amount of oxygen are unreliable due to variations in silicone species and adsorbates. The problem is not only academic, since silicates are not as harmful as silicones and are sometimes left on the surface after detergent cleaning (silicates are often included in the cleaning agent). Two techniques can help, SIMS and Fourier transformed infrared spectroscopy (FTIR). Even in low sensitivity SIMS, fragments such as $Si(CH_3)_3^+$, at 73 m/e can be detected and do not overlap in the spectrum with other intense hydrocarbon fragments. For FTIR, the sample can be rinsed with hexane, which is a good solvent for silicone oils and greases, as mentioned before. The resulting solution is deposited directly on the window used for the IR transmission measurements in a benchtop instrument and the data acquisition is carried out after evaporation of the hexane. The typical absorption features are in the 800−1300 $cm^{-1}$ region. The sensitivity of this elution method is potentially sub-monolayer and depends on the size of the rinsed surface. It can be applied with exactly the same method of elution to hydrocarbon contamination and can give an independent confirmation of the XPS results. The advantage with respect to XPS is the sampling, and therefore averaging capability, over a much larger surface area. For parts which hare difficult to transport, the test of elution can in principle be performed close to the surface to be tested and even the analysis can be done with portable FTIR instruments now available on the market. The method with elution is still superior in terms of quantitative reliability of the results compared to a direct measurement in reflection, also possible with a portable instrument. The reflection measurements on the real part are influenced by the substrate material (possibly due to the optical properties of the surface oxide layer) and roughness.

Thermal outgassing is another method of testing surface cleanliness, which probes the functional performance in static vacuum conditions. This is extensively discussed in another presentation at this CAS school [33].

## 3.2 Tests for dynamic vacuum performance

In the particular case of surfaces, which are designed to be exposed to a particle beam, as in the case of an accelerator beam pipe, the actual pressure or dynamic vacuum is determined by particle induced desorption. An overview of the different processes occurring is given in Refs. [34, 35]. For such cases it is safe to verify the cleanliness levels also by ESD. For the most common cleaning procedures used at CERN this has been done, as described in Ref. [15]. Often ESD is applied to tube-shaped samples measured after baking at 150°C to reduce the level of water in the residual gas, which would mask other fragments. The typical irradiation dose where the ESD yield decreases by a factor of 10 is about $10^{16}$ electrons/cm$^2$ ($10^{-3}$ Clb/cm$^2$) due to a progressive cleaning of the surface. This means that the dose must be limited in order to acquire relevant data for the characterization of the surface in an unconditioned state.

A similar technique is based on ion stimulated desorption [36]. An ion source in the KeV range can be obtained from usual SIMS or sputtering ion guns, where defocusing or scanning should be applied to reduce the current density and increase the irradiated surface. This will avoid surface cleaning during measurement on the one hand, and on the other hand, it will deliver a sufficiently high signal to noise. The main tendencies of ion stimulated desorption, in regard to desorbed species and intensities, correlate well with ESD [21]. A special case of ion stimulated desorption is represented by high energy (MeV/nucleon) highly ionized heavy ions [37]. This phenomenon is particularly important for ion storage rings where the impact of lost ions can induce pressure bursts. The lifetime of such particles is extremely sensitive to the residual gas pressure and has a positive feedback mechanism, since ions with a modified charge will diverge from the beam, impinge on the chamber wall, and desorb more gas. It has been shown that desorption coefficients are some orders of magnitude higher for such ions that, for instance, for electrons. The phenomenon is not completely understood yet, but the usual relation between surface contamination level and desorption yield is confirmed. For instance, coatings of the surface, such as getters or noble metals, which can be easily cleaned by baking in situ, exhibit lower yields than bare stainless steel surfaces.

To measure photon stimulated desorption (PSD), which is relevant in all the cases where synchrotron radiation impinges on the beam chamber walls, the only suitable source is generally obtained from synchrotron radiation itself at the necessary critical energy. Experiments along this line are performed [20, 34] by operating a large machine rather than a small set up for laboratory size experiments.

## 3.3 On-line and off-line quality control

The ideal quality control for a cleaning plant consists of the real time monitoring of the surface cleanliness immediately after processing. This would enable provision of the necessary bath maintenance in time and avoid delivery of parts which are not perfectly satisfactory. Some of the methods considered in the literature for the characterization of surface cleanliness are listed in Table 1. Unfortunately, none of these techniques can be applied as a fast selection test on line to a series of cleaned objects having different base material, shape, size, as in the case of parts cleaned in a facility for an accelerator. It is worth noting that the simplest fast monitoring for the cleanliness of treated parts is the observation of the wetting behaviour of the piece immediately after rinsing. The piece is considered clean if it remains covered by a uniform layer of water when it is lifted out of the rinsing bath or after spraying some water on it. This type of control relies on the experience of the operator of the cleaning station. Available commercial portable instruments for water contact angle measurement in situ have shown low reliability. In practice, the only safe method to avoid the installation of a contaminated part is an outgassing acceptance test for each component before installation [33]. A series of tests of samples is performed in parallel to avoid as much as possible the delivery of unsatisfactory parts.

### 3.4 Evaluation of the effectiveness of a cleaning procedure

The evaluation of the effectiveness of a cleaning method is verified by contaminating a sample with a well-known blend, cleaning, and analysing the surface for instance by XPS. A sufficiently large area and number of samples (4−5) should be measured, in order to average over statistical variations within the same cleaning run. A similar rationale is recommended also by international standards [38]. The mixture should be chosen to contain chemicals which are representative of a real contamination supposed to occur on the parts arriving at the cleaning plant. For instance, a mixture of oils and greases used during machining at the local workshop or vacuum pump oil which might be present on previously used parts, can be reasonable choices [30, 31]. At CERN such a method has been adopted also to assess the quality of cleaning procedures used by external manufacturers of parts to be inserted in UHV.

An especially elegant way of measuring the amount of residue left on the surface from a previous well-defined contamination is the method of the radioactive tracer [15]. The contamination molecules carry radioactive isotopes and the level of radioactivity after cleaning measures directly how effectively such a contamination has been removed over the all sample surface.

### 3.5 Packaging and storing cleaned parts

Recontamination occurs through adsorption of contaminants from air or through use of incorrect packaging methods. Upon air exposure, a clean metallic surface, for instance a sputter cleaned copper surface, forms a layer of oxide, then possibly part of it converts to hydroxide and in any case, it will be covered by adsorbed water and hydrocarbons. This occurs because the surface energy of an atomically clean metal or an oxide is some 10−100 times higher than that of water or hydrocarbons (1850 mJ/m$^2$ for clean metallic copper, around 800 mJ/m$^2$ for copper oxide [39], 72 mJ/m$^2$ for water, and 25 mJ/m$^2$ for alkanes). Hydrocarbons have the lowest surface energy and can cover such a surface in a dynamic process, which possibly results in a contamination layer including water and hydrocarbon molecules, the latter with the non-polar regions pointing toward air. Such a process results in the growth of carbon contamination, as illustrated in Fig. 8. Similar curves were measured by ellipsometry [27] on precision cleaned silicon wafers. Two further conclusions should be drawn from Fig. 8. First, large parts which will be inevitably exposed to air before packaging or tight closure will always exhibit non-vanishing hydrocarbon coverage on the surface. Second, comparison of the effectiveness of cleaning procedures is meaningful only when air exposure time and the storage method (see below) have been correctly defined in advance.

In order to avoid recontamination parts should be used as soon as possible after cleaning. This is not always possible in the case of construction of large plants where large series are cleaned, transported, and installed. Proper packaging to protect the part from contamination during storage is required. In Figure 9 [31] a comparison of simple storage and packaging methods are shown. All the samples were cleaned in the same run in a detergent bath, measured by XPS, and stored, wrapped in aluminium foil, in air (protected from dust in a Petri dish), in a polyethylene bag inside a pure-polyethylene bag, and inserted in a polyethylene bag after wrapping in aluminium foil. It is clear that inserting the samples into a polyethylene foil can have dramatic effects and moreover the result depends on the polyethylene quality. However, packaging in a polymer bag has obvious advantages of protection from macroscopic contamination during transport. Wrapping the parts in clean aluminium foil (in this case the common grade used for food packaging) before packaging them in the polyethylene bag removes completely the effect of the surrounding polymer bag. The sample remains as clean as if wrapped in aluminium foil alone. Such a method is obviously not suitable in cases where aluminium traces can provoke adverse effects on the part.

It is worth noting that contamination increases during the first month of storage, but its amount saturates and the further increase measured after 6 months exposure is moderate. This is due to the fact that initial adsorption on the high-energy surface is much more favoured than that on the contaminated low-energy surface.

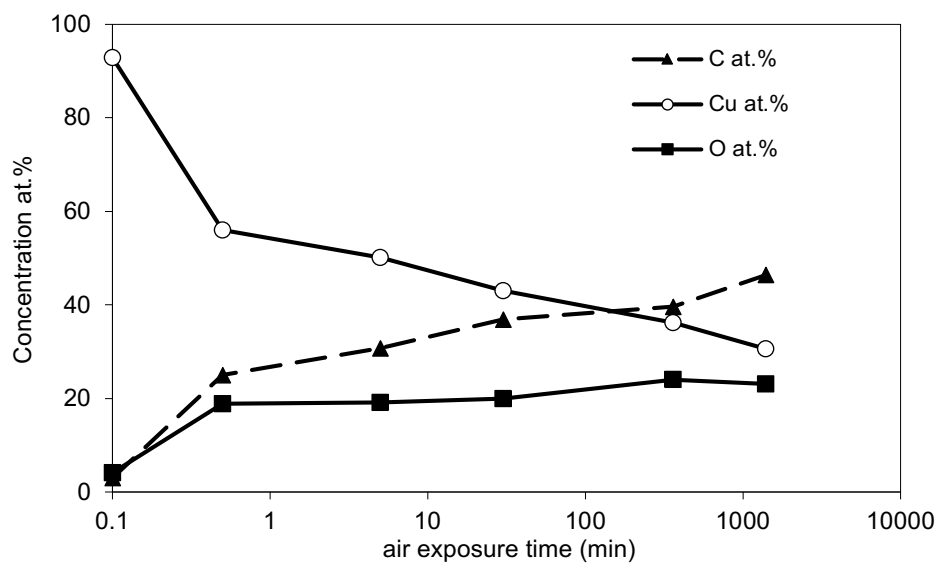

**Fig. 8:** Evolution for the composition, measured by XPS, of a sputter cleaned copper surface as a function of the air exposure time in laboratory.

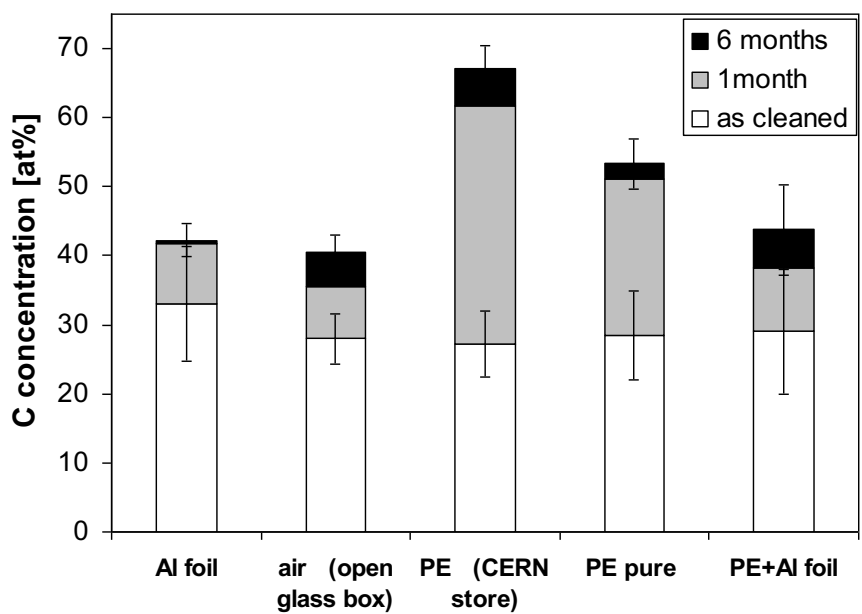

**Fig. 9:** Effect of recontamination of copper samples cleaned in a detergent bath and stored in different ways (average over four samples in each case).

A remark should be made concerning polyethylene bags. Such a packaging should only be considered as a mechanical barrier to dust and not as a barrier against humidity. The common polyethylene (typically 0.1 mm thick) bag filled with dry nitrogen does not protect contents from humidity for more than few hours, as can be easily tested by packaging silica gel or by calculation of the water diffusion with the literature permeation values for polyethylene.

## 4 Surface properties and contamination

### 4.1 Wetting and surface energy

In addition to the static and dynamic vacuum, other properties which are relevant for UHV application can be influenced by the surface cleanliness. During air exposure adsorbates such as hydrocarbon and water cover easily any clean surface and lower its surface energy. This prevents the adhesion of further coatings, for instance those deposited by magnetron sputtering or evaporation. This is the main reason for cleaning a surface before coating or even gluing. In principle, a single monolayer of tightly packed organic molecules covering the surface, such as a deposited self-assembled monolayer or a Langmuir Blodgett film, is sufficient to reduce drastically the surface energy [40]. In the case of hydrocarbons adsorbed in a disordered layer during air exposure, this occurs more gradually and a larger total thickness is needed. It has been shown [41] that the surface energy decreases down to the level of bulk hydrocarbons when the over layer reaches a thickness of 2−3 nm. In the same way the water contact angle increases [30], causing a decrease in wettability, as a function of air exposure time.

### 4.2 Secondary electron yield

The secondary electron yield (SEY) of the surfaces exposed to the particle beam is a further quantity which is relevant for particle accelerators. This quantity influences the beam stability and in the worst case can maintain the so called multipacting or resonant electron multiplication or electron cloud effect [42,43]. For common technological materials such as stainless steel, aluminium, and copper, the SEY of the air exposed surface is much higher than that for the clean metal or the corresponding oxide. As can be concluded from the shift of the maximum yield toward lower primary energies [44], the SEY increase is mainly due to coverage with a layer of contaminant, which has itself a high SEY. Since thin layers of water have been shown to be ineffective in causing such an increase, the origin of the fast-initial increase must be in the adsorbed airborne hydrocarbons.

For getter materials, the activation process removes the surface contamination by transforming the hydrocarbons in carbides and by letting the oxygen of the oxide diffuse into the bulk; simultaneously, the SEY is reduced [45]. For the more common metals used for accelerator vacuum chambers the SEY can be progressively reduced by irradiation with electrons or photons (synchrotron radiation). This process is usually called conditioning and consists of surface cleaning through particle stimulated desorption and de-hydrogenation of the adsorbed hydrocarbons [28, 46].

As already presented in Section 3.5, cleaned surfaces can be re-contaminated due to the use of improper storage materials or even by storage in air. The effect of such a recontamination on the SEY is shown in Fig. 10. SEY always increases with storage time for all methods of packaging used. The curves show that this adverse effect can be limited by selecting suitable storage conditions. For instance, wrapping the parts in aluminium foil before inserting them into a polymer bag improves the situation in comparison with packaging in the polymer bag alone. The results of the SEY behaviour (Fig. 10) correlate well with those found for the carbon contamination increase as a function of storage time (Fig. 9). This fact proves that the high SEY of air exposed surfaces is related to contamination from airborne hydrocarbons.

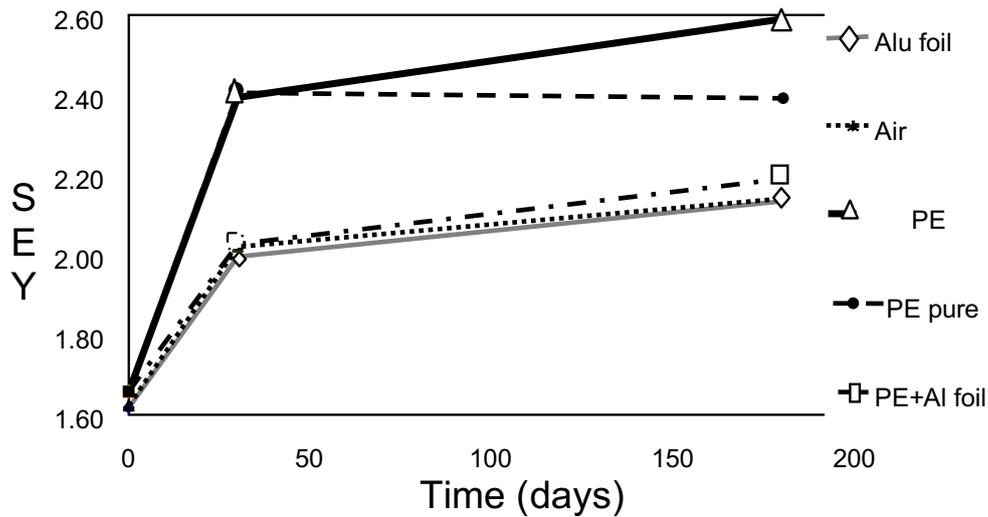

**Fig. 12:** Secondary electron yield of copper as a function of packaging method and storage time
(B. Henrist, N. Hilleret, C. Scheuerlein, M. Taborelli, unpublished).

## 5    Conclusions

The various methods of cleaning presented above are not a complete and universal recipe, but give some guidelines to optimize a cleaning method for a particular application. Even the most accurately selected procedure should be tested on the real parts to be cleaned or on a specimen which is representative of them in terms of shape, size, contamination, and surface composition. The possibility of cleaning the parts effectively should be considered and implemented from the initial stage of development, design, manufacturing, and the assembly process, in order to build 'cleanable' parts. A good rule is always to remember that cleaning is only necessary because at some stage there is contamination added to the parts. Typically, this occurs during the manufacturing process. Therefore, the safest way of proceeding is not to rely on cleaning, but to conceive a fabrication process which avoids the use of the most relevant contaminants.